\documentclass[12pt]{iopart}

\usepackage{amsfonts, amssymb, amsgen}
\usepackage{multicol}
\usepackage{braket}

\usepackage{cite}
\usepackage{bm}

\newcommand{\eq}[1]{\begin{equation} #1 \end{equation}}
\newcommand{\eqa}[2]{\begin{equation} #1 \label{#2} \end{equation}}
\newcommand{\balign}[1]{\begin{eqnarray} #1 \end{eqnarray}}

\newcommand{\mx}[1]{
\left(
\begin{array}{rr}
#1 \\
\end{array}
\right)
}

\newcommand{\todayd}{\the\year/\the\month/\the\day}

\newcommand{\bib}{\bibitem}

\newcommand{\up}{\uparrow}
\newcommand{\down}{\downarrow}
\newcommand{\lr}{\leftrightarrow}

\newcommand{\lb}{\label}
\newcommand{\nt}{\nonumber}
\newcommand{\eqref}[1]{(\ref{#1})}

\newcommand{\bel}{\begin{easylist}}
\newcommand{\eel}{\end{easylist}}
\def \({\left(}
\def \){\right)}
\def \[{\left[}
\def \]{\right]}

\newcommand{\sumtwo}[2]%
{\mathop{\sum_{#1}}_{#2}}
\newcommand{\sumthree}[3]%
{\mathop{\mathop{\sum_{#1}}_{#2}}_{#3}}
\newcommand{\sumfour}[4]%
{\mathop{\mathop{\mathop{\sum_{#1}}_{#2}}_{#3}}_{#4}} 
\newcommand{\prodtwo}[2]%
{\mathop{\prod_{#1}}_{#2}}
\newcommand{\mintwo}[2]%
{\mathop{\min_{#1}}_{#2}}
\newcommand{\maxtwo}[2]%
{\mathop{\max_{#1}}_{#2}}
\newcommand{\maxthree}[3]%
{\mathop{\mathop{\max_{#1}}_{#2}}_{#3}}
\newcommand{\limtwo}[2]%
{\mathop{\lim_{#1}}_{#2}}
\newcommand{\suptwo}[2]%
{\mathop{\sup_{#1}}_{#2}}
\newcommand{\supthree}[3]%
{\mathop{\mathop{\sup_{#1}}_{#2}}_{#3}}
\newcommand{\supfour}[4]%
{\mathop{\mathop{\mathop{\sup_{#1}}_{#2}}_{#3}}_{#4}} 
\newcommand{\inftwo}[2]%
{\mathop{\inf_{#1}}_{#2}}
\newcommand{\infthree}[3]%
{\mathop{\mathop{\inf_{#1}}_{#2}}_{#3}}
\newcommand{\inffour}[4]%
{\mathop{\mathop{\mathop{\inf_{#1}}_{#2}}_{#3}}_{#4}} 
\newcommand\calT{{\cal T}}

\newcommand{\bbZ}{\mathbb{Z}}
\def\rnum#1{\resizebox{0.5em}{\height}{\expandafter{\romannumeral #1}}}
\def\Rnum#1{\resizebox{0.5em}{\height}{\uppercase\expandafter{\romannumeral #1}}}

\begin{document}

\title[Connection between quantum-many-body scars and the AKLT model]{Connection between quantum-many-body scars and the AKLT model from the viewpoint of embedded Hamiltonians}

\author{Naoto Shiraishi}

\address{Department of physics, Gakushuin university, Toshima-ku, Tokyo, Japan}
\ead{naoto.shiraishi@gakushuin.ac.jp}
\vspace{10pt}

\begin{abstract}
We elucidate the deep connection between the PXP model, which is a standard model of quantum many-body scars, and the AKLT Hamiltonian.
Using the framework of embedded Hamiltonians, we establish the connection between the PXP Hamiltonian and the AKLT Hamiltonian, which clarifies the reason why the PXP Hamiltonian has nonthermal energy eigenstates similar to the AKLT state.
Through this analysis, we find that the presence of such nonthermal energy eigenstates reflects the symmetry in the AKLT Hamiltonian.
\end{abstract}

%\maketitle

\section{Introduction}

Triggered by elaborated experiments of cold atoms~\cite{Gri, Lan}, the presence or absence of thermalization (relaxation to the equilibrium state) in isolated quantum many-body systems has been intensively studied in theoretical physics.
Based on numerical results, most non-integrable many-body systems are considered to satisfy the eigenstate thermalization hypothesis (ETH), which claims that all energy eigenstates are thermal, that is, indistinguishable from the equilibrium state as long as we observe macroscopic quantities~\cite{Deu, Sre96, Rig08, KIH, BMH}.
Combining the above fact and the fact that the ETH is a sufficient condition for thermalization, the ETH is considered to be a key component to understand thermalization phenomena.
On the other hand, recent studies have revealed the fact that some non-integrable systems do not satisfy the ETH and have non-thermal energy eigenstates~\cite{SM, MS, Ber17, Mou, Shi17, Konik1, Konik2, KN}.
In particular, an experiment of Rydberg atoms reported unexpected long-lived oscillation which appears not to thermalize~\cite{exp}.
This experiment attracts the interest of many researchers, and various theoretical studies have tackled the intriguing phenomena.
Soon after the experiment, it has been pointed out that this Rydberg atom system can be mapped onto a spin-1/2 chain so-called PXP model~\cite{Tur}.
Analogous to the semiclassical chaotic billiard systems~\cite{scar}, this non-thermalizing behavior is called {\it quantum many-body scars}~\cite{Tur, Ho}.
To derive the scars in the PXP model, various theoretical investigations have been attempted including a perturbative approach~\cite{KLC, Choi}, matrix-product state (MPS) representations~\cite{Ho}, Floquet random unitary circuits~\cite{PP}, quantum topological phases~\cite{Ok}, extending the local Hilbert space for oscillation~\cite{BMP}, and the quasiparticle picture~\cite{ISX}.
See also a review paper~\cite{AABS}.
An important progress has been achieved by Lin and Motrunich~\cite{LM18}, where they explicitly construct two non-thermal energy eigenstates of the PXP model.
They also show that the non-thermal energy eigenstates have a connection to the Affleck-Kennedy-Lieb-Tasaki (AKLT) state~\cite{AKLT} through MPS representations.
However, at the Hamiltonian level, the connection between the PXP model and the AKLT Hamiltonian has not yet been clarified so far.
Since the AKLT system has been studied well~\cite{Aro, Ber17}, establishing the connection between the PXP model and the AKLT Hamiltonian will help deepen our understanding of quantum many-body scars.

In this paper, we present a full picture of connection between the PXP model and the AKLT Hamiltonian.
We employ the method of embedded Hamiltonians, which construct non-integrable Hamiltonians with desired states as its non-thermal energy eigenstates~\cite{SM}.
We demonstrate that the PXP model is indeed a variant of the AKLT Hamiltonian from the viewpoint of this embedding method.
The AKLT Hamiltonian is expressed as the sum of projection operators, and the PXP model is expressed as the sum of local Hamiltonians sandwiched by the same projection operators.
Through this analysis, we also find that the scarred eigenstates reflect the symmetry of the AKLT state.
Our result also elucidates the underlying structure of quantum many-body scars.
By taking this structure into account,  we can systematically construct various scarred systems.

This paper is organized as follows.
The first three sections are devoted to reviews of previous studies.
In \sref{s:AKLT}, we briefly review the AKLT model and its MPS representation.
In \sref{s:embed}, we briefly review the method of embedded Hamiltonian and see how the violation of the ETH is triggered.
In \sref{s:PXP}, we introduce the so-called PXP model~\cite{Tur}, which is a model of quantum many-body scars.
Following Lin and Motrunich~\cite{LM18}, we employ the block representation and show two exact non-thermal energy eigenstates of the PXP model.
In the next three sections, we investigate the connection between the AKLT Hamiltonian and the PXP model.
In \sref{s:analysis}, we rewrite the PXP Hamiltonian with keeping the embedding method in mind, and prove that the candidate of a non-thermal energy eigenstate is indeed an energy eigenstate in a direct and a little tedious way.
The connection between the AKLT Hamiltonian and the PXP Hamiltonian is demonstrated in \sref{s:PXP-embed} through the method of embedded Hamiltonian.
In \sref{s:sym}, relying this connection we provide a more transparent proof of existence of a non-thermal energy eigenstate in the PXP model.
We reveal that the scar of the PXP model reflects the frustration-free structure and spatial symmetry of the AKLT Hamiltonian.

\section{AKLT model}\lb{s:AKLT}

We first briefly review the AKLT model~\cite{AKLT}.
Let $\ket{1}$, $\ket{0}$, $\ket{-1}$ be the three eigenstates of the spin operator $S_z$ with eigenvalues 1, 0, $-1$, respectively.
The AKLT Hamiltonian is essentially equivalent to the sum of the projection operator $P^{S=2}$ acting on the two neighboring sites:
\eq{
H^{\rm AKLT}=\sum_i P_{i,i+1}^{S=2},
}
where the projection operator $P^{S=2}$ projects onto the subspace spanned by the following five states:
\balign{
\ket{\psi_1}&=\frac 1{\sqrt{2}} (\ket{0,-1}+\ket{-1,0}) \\
\ket{\psi_2}&=\frac 1{\sqrt{2}} (\ket{0,1}+\ket{1,0}) \\
\ket{\psi_3}&=\frac1{\sqrt{6}} (\ket{1,-1}+\ket{-1,1}+2\ket{0,0}) \\
\ket{\psi_4}&=\ket{1,1} \\
\ket{\psi_5}&=\ket{-1,-1}.
}
This subspace is also characterized by the total spin $S=2$.

The AKLT state is also described by the following matrix-product state (MPS) representation:
\eq{
A^{1}=\sqrt{\frac 23}\mx{0&1 \\0&0}, \hspace{12pt} A^{0}=-\sqrt{\frac 13}\mx{1&0 \\0&-1}, \hspace{12pt} A^{-1}=-\sqrt{\frac 23}\mx{0&0 \\1&0}.
}
The AKLT state satisfies
\eq{
P_{i,i+1}^{S=2}\ket{\rm AKLT}=0
}
for any $i$, and hence is a ground state of $H^{\rm AKLT}$.
It has been established that the AKLT state is the unique gapped ground state of the AKLT Hamiltonian.

\section{Embedded Hamiltonian}\lb{s:embed}

We here briefly review the method of embedded Hamiltonians~\cite{SM}.
Let $\calT$ be a subspace of states which we want to embed as energy eigenstates.
We introduce a set of local projection operators $\{ P_i\}$, where $P_i$ acts on sites near the site $i$ and satisfies
\eq{
P_i\ket{\psi}=0
}
for any state $\ket{\psi}\in \calT$.
We construct an embedded Hamiltonian as
\eq{
H=\sum_i P_i h_i P_i +H',
}
where $h_i$ is an arbitrary local Hamiltonian acting on sites near the site $i$, and $H'$ is a Hamiltonian satisfying $H'\ket{\psi}\in \calT$ if $\ket{\psi}\in \calT$.
Then the Hamiltonian $H$ has a set of energy eigenstates which is a basis of the subspace $\calT$.
In particular, if $\calT$ consists of a single state $\ket{\Psi}$, this state $\ket{\Psi}$ is an energy eigenstate at finite temperature.
We emphasize that $h_i$ is arbitrary, and thus the constructed Hamiltonian is in general highly complicated and non-integrable.

In spite of the simplicity of construction, the physical implication from such Hamiltonians is not negligible.
This method reveals the existence of infinitely many non-integrable Hamiltonians with non-thermal energy eigenstate at finite temperature (i.e., violation of the ETH).
Moreover, using this method we design non-integrable and chaotic Hamiltonians which have arbitrary desired states as their energy eigenstates.

\section{Quantum many-body scars and PXP model}\lb{s:PXP}

The quantum many-body scar was first found in the experiment of Rydberg atoms, where we can observe non-thermalizing oscillation~\cite{exp}.
The dynamics of this system is well described by an $S=1/2$ chain, the so-called PXP model~\cite{Tur}:
\eq{
H^{PXP}=\sum_{i=1}^L P_{i-1}X_iP_{i+1},
}
where $P:=\ket{\down}\bra{\down}$ is the projection operator onto the down state and $X:=\ket{\up}\bra{\down}+\ket{\down}\bra{\up}$ is the Pauli matrix.
For a technical reason explained later, the system size $L$ is supposed to be an even number.
To avoid confusion with the symbols in the AKLT model, we denote the ground state and the excited state of a single Rydberg atom by $\ket{\down}$ and $\ket{\up}$, not by conventional symbols $\ket{0}$ and $\ket{1}$.
We employ the periodic boundary condition for convenience.

Owing to the projection operators, two neighboring up states $\ket{\up \up}$ never flip.
Since we are not interested in these {\it frozen} blocks, we restrict the state space into its subspace free from two neighboring up states.
In the remainder of this paper, our analysis is performed in this subspace.

Following the Lin-Motrunich paper~\cite{LM18}, we use the block representation of states: 
We regard two neighboring odd and even sites as a single block site (i.e., (1,2), (3,4), (5,6),... are regarded as blocks).
Since the system size $L$ is assumed to be even, the block representation is consistent.
Three possible states of a block site $\ket{\down \down}$, $\ket{\up \down}$, and $\ket{\down \up}$ are denoted by $\ket{O}, \ket{L}, \ket{R}$, respectively.
Eq.(10) in the Lin-Motrunich paper~\cite{LM18} shows that the PXP Hamiltonian is also expressed as
\balign{
H^{PXP}=&\sum_{b=1}^{L/2} h_{b,b+1}, \\ 
h_{b,b+1}=&(\ket{R}\bra{O}+\ket{O}\bra{R})_b \otimes (I-\ket{L}\bra{L})_{b+1} \nt \\
&+(I-\ket{R}\bra{R})_b \otimes (\ket{L}\bra{O}+\ket{O}\bra{L})_{b+1},
}
where $b\in \{ 1,2,\ldots , L/2\}$ is the label of a block which consists of odd and even sites with $2b-1$ and $2b$.
In addition, they show that $\ket{\Phi_1}$ (given in (4) and (12) in Ref.~\cite{LM18}), whose (unnormalized) MPS representation is
\eq{
A^O=\mx{0&-1 \\ 1&0}, \hspace{12pt} A^R=\mx{\sqrt{2}&0 \\0&0}, \hspace{12pt} A^L=\mx{0&0 \\0&-\sqrt{2}},
}
is an energy eigenstate of $H^{PXP}$ with zero energy.
This clearly shows the violation of the ETH in the PXP model.
(We, however, note that the possibility that the PXP Hamiltonian is integrable is not excluded.
A technique invented recently~\cite{Shi18} may help to exclude this unwanted possibility.)
By considering another block representation where we regard two neighboring even and odd sites as a single block site (i.e., (2,3), (4,5), (6,7),... are regarded as blocks), we obtain another non-thermal energy eigenstate with zero energy $\ket{\Phi_2}$.
These two states, $\ket{\Phi_1}$ and $\ket{\Phi_2}$, provide the scar of $\mathbb{Z} _2$ synchronization in this PXP model.

Here we can see the strong similarity to the AKLT state.
In fact, Lin and Motrunich have pointed out that the AKLT state $\ket{\rm AKLT}$ and the state $\ket{\Phi_1}$ are identical by adopting the correspondence $O\lr 0$, $L\lr -1$, and $R\lr 1$ on the odd site and $O\lr 0$, $L\lr 1$, and $R\lr -1$ on the even site.
In other words, the energy eigenstate $\ket{\Phi_1}$ and the AKLT state are equivalent under a proper gauge transformation.
We call the gauges corresponding these two states as {\it PXP gauge} and {\it AKLT gauge}, respectively.

However, at the level of Hamiltonians, the connection between the AKLT Hamiltonian and the PXP Hamiltonian has still been unclear.
In the remainder of this paper, we address this problem.

\section{Analysis of PXP model}\lb{s:analysis}

\subsection{Rewriting PXP Hamiltonian}

Based on this MPS representation, we introduce a projection operator $P^{PXP}$, which is equivalent to the $P^{S=2}$ under the gauge transformation.
This projection operator projects onto the subspace spanned by
\balign{
\ket{\phi_1}&=\frac 1{\sqrt{2}} (\ket{OR}+\ket{LO}) \\
\ket{\phi_2}&=\frac 1{\sqrt{2}} (\ket{OL}+\ket{RO}) \\
\ket{\phi_3}&=\frac1{\sqrt{6}} (\ket{RR}+\ket{LL}+2\ket{OO}) \\
\ket{\phi_4}&=\ket{RL} \\
\ket{\phi_5}&=\ket{LR}.
}
Due to the correspondence between the AKLT state and $\ket{\Phi_1}$, we find
\eq{
P_{b,b+1}^{PXP}\ket{\Phi_1}=0
}
for any $b$.

We rewrite the Hamiltonian $H^{PXP}$ as~\cite{LM18}
\balign{
H^{PXP}=&\sum_b \[ -h^2_{b,b+1}+\frac 12 h^1_{b,b+1}\] , \\
h^2_{b,b+1}:=&(\ket{R}\bra{O}+\ket{O}\bra{R})_b \otimes (\ket{L}\bra{L})_{b+1} \nt \\
&+(\ket{R}\bra{R})_b \otimes (\ket{L}\bra{O}+\ket{O}\bra{L})_{b+1},  \\
h^1_{b,b+1}:=& (\ket{R}\bra{O}+\ket{O}\bra{R}+\ket{L}\bra{O}+\ket{O}\bra{L})_b \nt \\
&+(\ket{R}\bra{O}+\ket{O}\bra{R}+\ket{L}\bra{O}+\ket{O}\bra{L})_{b+1}, \lb{h'}
}
where we have
\eq{
[h^1_{b,b+1}, P_{b,b+1}^{PXP}]=0, \lb{commute-1} 
}
\eq{
[h^2_{b,b+1}, P_{b,b+1}^{PXP}]=0. \lb{commute-2} 
}
In other words, two Hamiltonians $h^1_{b.b+1}$ and $h^2_{b,b+1}$ keep the subspace of the Hilbert space associated with $P_{b,b+1}^{PXP}=1$ and that with $P_{b,b+1}^{PXP}=0$ invariant, and thus we can treat the action of these Hamiltonians in each subspace separately.
The commutation relations \eqref{commute-1} and \eqref{commute-2} can be confirmed directly as
\balign{
h^2 \ket{\phi_2}&=\sqrt{2} \ket{\phi_4} \\
h^2 \ket{\phi_4}&=\sqrt{2} \ket{\phi_2} \\
h^2 \ket{\phi_i}&=0 \ \ (i=1,3,5) \\
& \nonumber \\
h^1 \ket{\phi_1}&=\sqrt{2}\ket{\phi_5}+\sqrt{3} \ket{\phi_3} \\
h^1 \ket{\phi_2}&=\sqrt{2}\ket{\phi_4}+\sqrt{3} \ket{\phi_3} \\
h^1 \ket{\phi_3}&=\sqrt{3} (\ket{\phi_1}+\ket{\phi_2}) \\
h^1 \ket{\phi_4}&=\sqrt{2} \ket{\phi_2} \\
h^1 \ket{\phi_5}&=\sqrt{2} \ket{\phi_1}.
}

\subsection{Proving $h^2_{b,b+1}\ket{\Phi_1}=0$}

It is easy to see that $h^2$ only permutes some states in the subspace with $P_{b,b+1}^{PXP}=1$ and satisfies $h^2\ket{\eta}=0$ for any $\ket{\eta}$ in the subspace with $P_{b,b+1}^{PXP}=0$.
With recalling the relation $P_{b,b+1}^{PXP}\ket{\Phi_1}=0$ for any $b$, we conclude that
\eq{
h^2_{b,b+1}\ket{\Phi_1}=0
}
for any $b$.

\subsection{Proving $\sum_b h^1_{b,b+1}\ket{\Phi_1}=0$}\lb{pf-h1}

We now consider how $h^1$ acts on $\ket{\Phi_1}$.
To say a result, 
\eqa{
\sum_b h^1_{b,b+1}\ket{\Phi_1}=0
}{h1=0}
is satisfied.
We note that an individual $h^1$ does not have this property (i.e., $h^1_{b,b+1}\ket{\Phi_1}\neq 0$).

We first introduce the rotated basis states of a single site:
\balign{
\ket{+}&:=\frac{1}{\sqrt{2}}(\ket{R}+\ket{L}), \\
\ket{-}&:=\frac{1}{\sqrt{2}}(\ket{R}-\ket{L}).
}
Using these states, the basis of the subspace associated with $1-P^{PXP}$ is expressed as
\balign{
\ket{\phi_6}&=\frac{1}{\sqrt{2}}(\ket{O+}-\ket{+O}) \\
\ket{\phi_7}&=\frac{1}{\sqrt{2}}(\ket{O-}+\ket{-O}) \\
\ket{\phi_8}&=\frac{1}{\sqrt{3}}(\ket{OO}-\ket{++}-\ket{--}) \\
\ket{\phi_9}&=\frac{1}{\sqrt{2}}(\ket{+-}+\ket{-+}) 
}
and the MPS representation of $\ket{\Phi_1}$ is written as
\eq{
A^O=\mx{0 &-1 \\ 1 & 0}, \ \  A^+=\mx{1&0 \\ 0&-1}, \ \ A^-=\mx{1&0 \\ 0&1}.
}
In this subspace, $h^1$ acts as
\eq{
\ket{\phi_7}\bra{\phi_9}+\ket{\phi_9}\bra{\phi_7}.
}
Since $\ket{\cdots O-\cdots}$ and $\ket{\cdots -O\cdots}$ (and $\ket{\cdots O+\cdots}$ and $\ket{\cdots +O\cdots}$) have the same coefficients in $\ket{\Phi_1}$ owing to the above MPS representation, as far as $\ket{\Phi_1}$ we can regard $h^1$ as a flip operator of $\ket{+}$ and $\ket{O}$ next to $\ket{-}$.

We now show that all the coefficients of $\sum_b h^1_{b,b+1}\ket{\Phi_1}$ with computational basis states are zero.
In these basis states, we decompose spins into the islands of $-$ states and those of non-$-$ states (i.e., $+$ and $O$ states).
For example, a basis state $\ket{+---O+--+++O--}$ is decomposed into three $-$ islands and three non-$-$ islands as $[+] \ [---] \ [O+] \ [--] \ [+++O] \ [--]$.
The operator $h^1$ flips $+$ and $O$ at the edge of a non-$-$ island.

We fix a computational basis state, and focus on a $-$-island and two neighboring $+$ or $O$ states, $\ket{\cdots X--\cdots - Y\cdots}$, where $X,Y$ take $+$ or $O$, and spins between $X$ and $Y$ are $-$ states.
We consider the situation that $h^1$ acts on a state and the resulting state turns into $\ket{\cdots X--\cdots - Y\cdots}$.
Let $\bar{X}$ and $\bar{Y}$ be the opposite of $X$ and $Y$ in terms of $\{ +, O\}$.
As far as considering the region near this $-$ island, there are two possible states {\it before} operating  $\sum_b h^1_{b,b+1}$, that is, $\ket{\cdots \bar{X}--\cdots - Y\cdots}$ and $\ket{\cdots X--\cdots - \bar{Y}\cdots}$.
The coefficient of the state $\ket{\cdots X--\cdots - Y\cdots}$ after operating $h^1$s is equal to the sum of coefficients of $\ket{\cdots \bar{X}--\cdots - Y\cdots}$ and $\ket{\cdots X--\cdots - \bar{Y}\cdots}$ in $\ket{\Phi_1}$.
With noting that $A^-$ is the identity operator, and $A^+$ and $A^O$ satisfy
\eq{
A^+A^+=-A^OA^O, \ \ A^+A^O=-A^OA^+,
}
we find that the coefficients of $\ket{\cdots \bar{X}--\cdots - Y\cdots}$ and $\ket{\cdots X--\cdots - \bar{Y}\cdots}$ take the same absolute value with opposite signs.
This fact holds for any $-$ islands and any computational basis states, which directly implies that all the coefficients of computational basis states in $\sum_b h^1_{b,b+1}\ket{\Phi_1}$ are canceled.
This is the desired result \eqref{h1=0}.

\section{Representation as embedded Hamiltonian}\lb{s:PXP-embed}

In summary, the PXP Hamiltonian can be interpreted as an embedded Hamiltonian:
\eq{
H^{PXP}=\sum_b P_{b,b+1}^{PXP}H_bP_{b,b+1}^{PXP}+H',
}
where
\balign{
H_b:=&2\sqrt{2}(\ket{\phi_4}\bra{\phi_2}+\ket{\phi_2}\bra{\phi_4})_{b,b+1}+\sqrt{2} (\ket{\phi_1}\bra{\phi_5}+\ket{\phi_5}\bra{\phi_1})_{b,b+1} \nt \\
&+\sqrt{3}\[ \ket{\phi_3}(\bra{\phi_1}+\bra{\phi_2}) +(\ket{\phi_1}+\ket{\phi_2}) \bra{\phi_3}\] _{b,b+1},
}
and
\eq{
H'=\sum_b \ket{\phi_7}\bra{\phi_9}_{b,b+1}+\ket{\phi_9}\bra{\phi_7}_{b,b+1}.
}
The condition that $H'$ keeps $\ket{\Phi_1}$ is demonstrated in Sec.~\ref{pf-h1}.
It is worth comparing this Hamiltonian to the corresponding AKLT Hamiltonian in the PXP gauge:
\eq{
H^{\rm AKLT}=\sum_b P_{b,b+1}^{PXP}.
}
By construction of the embedded Hamiltonian, both $H^{PXP}$ and $H^{\rm AKLT}$ have the same eigenstate $\ket{\Phi_1}$ with zero energy, and in the PXP Hamiltonian this eigenstate $\ket{\Phi_1}$ is a nonthermal energy eigenstate, or a exact scarred state.

This structure reveals the following important fact:
By replacing $H_b$ to another arbitrary Hamiltonian $\tilde{H}_b$ and multiplying an arbitrary coefficient $k$ to $H'$ in the PXP Hamiltonian, the obtained Hamiltonian
\eq{
\tilde{H}_{PXP}=\sum_b P_{b,b+1}^{PXP}\tilde{H}_bP_{b,b+1}^{PXP}+kH'
}
still has the eigenstate $\ket{\Phi_1}$ as a non-thermal energy eigenstate.
Using this method, we can construct infinitely many scarred Hamiltonians.

\section{Symmetry in the AKLT model}\lb{s:sym}

We have shown that $\ket{\Phi_1}$ is a zero energy eigenstate of the PXP Hamiltonian by demonstrating $h^2_{b,b+1}\ket{\Phi_1}=0$ and $\sum_b h^1_{b,b+1}\ket{\Phi_1}=0$.
The former relation directly follows from the fact that $h^2_{b,b+1}$ is expressed in the form of $P^{PXP}_{b,b+1}hP^{PXP}_{b,b+1}$, while the latter relation is not obvious in the present form.
To clarify why the latter relations holds, we go back to the original AKLT model.
%In the PXP model, $H'$, or essentially the same as $\sum_b h^1_{b,b+1}$, has the eigenstate $\ket{\Phi_1}$ with zero eigenvalue, which is crucial in our construction.

We shall write down the operator $\sum_b h^1_{b,b+1}$ in the AKLT gauge.
By defining
\eq{
\ket{a}:=\frac{1}{\sqrt{2}}(\ket{1}+\ket{-1}),
}
$\sum_b h^1_{b,b+1}$ in the AKLT gauge is given by 
\eq{
\sum_i (\ket{a}\bra{0}+\ket{0}\bra{a})_i =\sum_i S^x_i,
}
where $S^x_i$ is the spin $x$ operator in the spin-1 system.
Due to this fact, in contrast to the case of $h^1$ (in the PXP gauge), it is easy to confirm the relation 
\eq{
\sum_i (\ket{a}\bra{0}+\ket{0}\bra{a})_i\ket{\rm AKLT}=\sum_i S^x_i \ket{\rm AKLT}=0,
}
which is a direct consequence of the symmetry of the AKLT Hamiltonian.
(More precisely, we used the absence of degeneracy in the ground state, the spatial symmetry of the AKLT Hamiltonian, which leads to $\braket{{\rm AKLT}|\sum_i S^x_i|{\rm AKLT}}=0$, and the commutation relation $[\sum_i S^x_i, H^{\rm AKLT}]=0$.)

In other words, the exact scars in the PXP Hamiltonian reflect the frustration-free structure and the $x$ symmetry of the AKLT Hamiltonian.
In the AKLT gauge, the AKLT Hamiltonian is expressed as
\eq{
H^{\rm AKLT}=\sum_i P^{S=2}_{i,i+1},
}
and the PXP Hamiltonian is expressed as
\eq{
H^{PXP}=\sum_i P^{S=2}_{i,i+1}H_iP^{S=2}_{i,i+1}+\sum_i S_i^x,
}
where $H_i$ is a proper local Hamiltonian.
Since the symmetry in the AKLT state is not restricted to the $x$ direction, we can construct other families of scarred Hamiltonians by employing other symmetries.

\section{Conclusion}

We have drawn the full picture of the connection between the PXP model and the AKLT model.
The AKLT Hamiltonian is a sum of projection operators $\sum_i P^{S=2}_{i,i+1}$, while the PXP Hamiltonian (in the AKLT gauge) is a sum of local Hamiltonians sandwiched by the same projection operators $\sum_i P^{S=2}_{i,i+1}H_iP^{S=2}_{i,i+1}$ with adding an $x$ magnetic field $\sum_i S_i^x$.
By construction, the first sum of the PXP Hamiltonian still has the AKLT state as its energy eigenstate with zero energy, and owing to the symmetry of the AKLT state the $x$ magnetic field does not disturb the AKLT state.
Our result confirms that the previously pointed connection between the scarred eigenstate $\ket{\Phi_1}$ in the PXP model and the AKLT state is not a coincidence but reflects the frustration-free structure of the AKLT Hamiltonian and the $x$ symmetry of the AKLT state.

We should comment some limitations of our results.
First, we assumed that the system size is assumed to be an even number in order to employ the block representation consistently.
The block representation with two sites successfully captures $\bbZ_2$ synchronization of the PXP model.
However, a system with odd length is also expected to show some scarred states and $\bbZ_k$ synchronization with $k\geq 3$~\cite{Tur}.
This observation is not explained by our approach at present.
Second, related to the above point, under a fixed block representation we find just one non-thermal energy eigenstate $\ket{\Phi_1}$.
On $\ket{\Phi_2}$, both $h^1$ and $h^2$ show highly nontrivial actions while they miraculously cancel, which is not characterized by the present representation.
As a result, we cannot describe two non-thermal energy eigenstates $\ket{\Phi_1}$ and $\ket{\Phi_2}$ with a single block representation at the same time.
To treat these two non-thermal energy eigenstates equally, we need a more sophisticated representation.

Various known properties of the AKLT model has already been established.
Applying them to the PXP model will lead to our deeper understanding of the PXP model and quantum many-body scars, which merits further investigation.

\ack
The author thanks Takashi Mori, Keiji Saito, and Sho Sugiura for fruitful discussion.
The author was supported by JSPS Grants-in-Aid for Scientific Research Grant Number JP19K14615.

\end{document}